# Interaction and transformation of metastable defects in intercalation materials

O. Yu. Gorobtsov[1], H. Hirsh[2], M. Zhang[2], D. Sheyfer[3], S. D. Matson[1], D. Weinstock[1], R. Bouck[1], Z. Wang[1], W. Cha[3], J. Maser[3], R. Harder[3], Y. Sh. Meng[2], A. Singer[1]

*1 – Materials Science and Engineering Department, Cornell University, Ithaca, NY 14853, USA*
*2 – Department of NanoEngineering, University of California, San Diego, La Jolla, California, 92093, USA*
*3 – Advanced Photon Source, Argonne National Laboratory, Argonne, Illinois 60439, USA*

Non-equilibrium defects often dictate macroscopic functional properties of materials. In intercalation hosts, widely used in rechargeable batteries, high-dimensional defects largely define reversibility and kinetics[1,2,3,4]. However, transient defects briefly appearing during ionic transport have been challenging to capture, limiting the understanding of their life cycle and impact. Here, we overcome this challenge and track operando the interaction and impact of metastable defects within $Na_xNi_{1-x}Mn_yO_2$ intercalation hosts in a charging sodium-ion battery. Three-dimensional coherent X-ray imaging[3,4,5] reveals transformation and self-healing of a metastable domain boundary, glissile dislocation loop, and stacking fault. A local strain gradient suggests a quantifiable difference in ion diffusion, coincident with the macroscopic change in diffusion coefficient. Analysis of the unexpected[4,6] defect anisotropy highlights the importance of mesostructure, suggesting a possible control approach and disputing the rigidity of the framework layers. The shared nature of oxygen framework layers makes our results applicable to a wide range of intercalation materials.

William Shockley summarized the importance of point defects: "Transistor electronics exist because of the controlled presence of imperfections in otherwise nearly perfect crystals"[7]. Point defects are equilibrium effects that enable sophisticated control routes of electronic properties. High-dimensional crystal defects – dislocations and domain boundaries – are intrinsically out-of-equilibrium phenomena, more challenging to understand and control than point defects. Nevertheless, successful models of dislocation nucleation, migration, and mutual interaction have transformed the understanding of plasticity and diffusion in metals[8]. While dislocations are rare in bulk ceramic crystals because of their high Peierls stress[8], recent research shows an abundance of dislocations[4,6,9,10] and domain boundaries[11,12,13] in nanosized ceramics. In layered oxides such as lithium-ion battery (LIB) and sodium-ion battery (SIB) electrode materials, these defects lead to capacity fade[4] or, in theory, facilitate ionic diffusion[13,14]. In SIBs – rapidly evolving safe[15], sustainable, and inexpensive energy storage solution for large-scale applications[16] – the cathode materials display particularly drastic structural rearrangements, which may negatively affect reversibility through defect formation[1,2].

Battery cathodes are complex, ceramics-polymer binder-carbon black composites, and understanding processes behind defect formation and dynamics requires operando measurements in multicomponent devices. The primary challenges lie in resolving defects, operando and at the nanoscale, in the electrochemically active material, and in precisely measuring the defect orientation in 3D, which is required in layered materials because of their anisotropy. Most conventional in-situ or operando techniques applied to battery materials[17] fail to accomplish both simultaneously. The multitude of traditional X-ray diffraction and absorption methods that allow operando and in-situ measurements[17] do not directly image defects in nanomaterials. In-situ electron microscopy[18] allows direct 2D visualisation of dislocations but is challenging to perform operando in multicomponent, multiparticle devices due to the absorption and radiation sensitivity of the battery components. Operando optical methods[19] do not have sufficient resolution. Operando neutron scattering[20] does not provide nano resolution and direct defect imaging.

To overcome these challenges and track operando defect formation and dynamics of cathode material in SIB, we applied synchrotron-based operando Bragg Coherent Diffractive Imaging (BCDI) (Methods, Supplementary Figure 7), applied previously to LIBs[3,4]. Operando BCDI combines three critical advantages over traditional characterization techniques: 1) coherent diffraction provides a 3D distribution of lattice displacement and atomic defect configuration with 10 nm resolution within individual nanoparticles; 2) X-ray penetration depth enables measurements during electrochemical cycling in a fully operational battery; 3) high X-ray flux allows capturing phenomena with a 1 minute resolution. Defects directly observed with BCDI include dislocations and dislocation loops[21] with a Burgers vector **b** component along the scattering vector **q**. These defects produce readily recognisable vortices in reconstructed displacement maps (Fig. 1, a, b)[3,5]. The direction of **b** determines the type of the observed dislocation: either perpendicular (edge, Fig. 1, a) or parallel (screw, Fig. 1, b) to the dislocation line. Observable planar defects, such as antiphase/out-of-phase domain boundaries, generate a lattice displacement in the domain volume in the direction of the momentum transfer **q** (Fig. 1 (c)). Among other planar defects, twin boundaries require multiple Bragg reflections[22] due to a different Bragg condition for the twins, and stacking faults in atomic crystals can generate a 2D-manifold of increased strain.

We imaged the structural evolution of individual sodium transition metal oxide (TMO) cathode nanoparticles of two compositions, P2-$Na_{0.78}Ni_{0.23}Mn_{0.69}O_2$[23] and P2-$Na_{0.66}Ni_{0.33}Mn_{0.66}O_2$[24], where the latter has a phase-change during electrochemical charge. Operando X-ray imaging reveals particles of both materials with a plate-like shape (example for P2-$Na_{0.78}Ni_{0.23}Mn_{0.69}O_2$ in Fig. 1, d) and a diameter within 100-1000 nm, in agreement with the ex-situ scanning electron microscopy images (SEM) of the material

(Supplementary Note 1). In a representative particle shown in Fig. 1, d, we observe a dislocation pair (magenta) formation process at a domain boundary (green) during charging. The defect signatures are evident in the two-dimensional cross-sections of the displacement within the particle (Fig. 2). A displacement domain (Fig 2, b) with a boundary perpendicular to the layer direction develops (Supplementary Note 3, Supplementary Video 1) and resolves into two vortices with opposite handedness (Fig 2, c, Supplementary Video 2). Dislocation lines are parallel to the **q** direction along (002), perpendicular to the TMO layers. The data reveals continuous displacement vortices with $\boldsymbol{b} \cdot \boldsymbol{q} = 2\pi$, meaning that $\boldsymbol{b}$ and the dislocation lines are aligned and the dislocations are predominantly screw.

The BCDI data allows us to develop a model for the mechanism of the dislocation pair nucleation (Fig 2, d). First, an out-of-phase domain boundary with constant displacement difference forms. Then, the boundary collapses to reconnect the layers, creating an antiparallel screw dislocation pair that reaches the particle surfaces, equivalent to a glissile dislocation loop ($\boldsymbol{b}$ in the plane of the loop) (Fig. 2, d, right). For nucleation of a dislocation pair in the P2-Na$_{0.78}$Ni$_{0.23}$Mn$_{0.69}$O$_2$ particle, the activation energy $E_c$ is estimated as $E_c \approx 1.1 \cdot 10^{-8} \frac{J}{m}$; the corresponding minimum shear stress assuming homogeneous nucleation would be $\sigma_{zp} \approx 80$ MPa (Methods, Supplementary Note 2). The high estimated stress coupled to the dislocation pair formation suggests heterogeneous nucleation: in the intermediate metastable configuration, the boundary stores energy and reduces the critical shear stress for creating the dislocation pair. After formation, a loop with oppositely oriented screw dislocations, removed far from the surface, would collapse in the absence of external stress because of the line tension. Instead, we find the opposite: the dislocations drift apart (Fig. 1, d) towards the surface of the particle. The dislocation loop expands, and after 30 mAHr/g (2 hours at C/10 charge rate), the screw dislocation pair reaches the particle surface. The expansion of the loop suggests continuous external stress perpendicular to the layers over $\sigma_{zp} > 20$ MPa (Methods, Supplementary Note 2).

Notably, in our loop formation model, the dislocation Burgers vector is half of the P2 unit cell constant. Therefore, the oxygen layers reconnect imperfectly (AB to BA, Delmas notation) within the loop (Fig. 2, d, inset). Akin to a stacking fault between a dissociated dislocation in metals, the imperfect connection can introduce additional strain. Indeed, as the battery charges (Fig 3, a-d), a planar strain signature oriented along the [001] direction appears at the position of the dislocation loop (Fig. 3, f). In a separate effect, before the dislocation loop nucleates, we observe a region of relatively lower interlayer distance (a compressive strain of ~-1*10$^{-4}$) when the domain boundary is present (Fig. 3, c, e). We identify two phenomena as the most plausible explanations for a local strain difference near a boundary: 1) coherency strain from the boundary itself[25] or 2) a relatively higher Na concentration caused by slower Na extraction near the boundary[14]. Because the local strain is concentrated in one (central) region, not along the whole domain boundary, the first option does not offer a complete explanation, leaving Na concentration as the more plausible strain origin. The relatively higher Na concentration in the presence of the domain boundary suggests that the boundary likely impedes ion diffusion. Our electrochemical impedance spectroscopy (EIS) measurements demonstrate that the ion diffusion dramatically slows down by approximately an order of magnitude in the beginning stages of P2-Na$_{0.78}$Ni$_{0.23}$Mn$_{0.69}$O$_2$ charge, correlating with the growing number of defects (see Methods and Supplementary Figure 3 for EIS details).

We established defect statistics by performing statistical analysis of operando X-ray diffraction over a large ensemble of individual cathode particles[26] (Fig. 4, a-c). The measured changes in the 3D diffraction intensity around a Bragg peak offer insight into misalignments and phase shifts in individual particles[27] unlike the average obtained in powder X-ray diffraction. Figure 4, a shows the typical evolution of a projected diffraction intensity of a P2-Na$_{0.78}$Ni$_{0.23}$Mn$_{0.69}$O$_2$ particle. For most particles, peak broadening along **q** is limited to 10-20%, while peak width perpendicular to **q** increases by over 100% (Fig 4, b), supporting preferential defect formation perpendicular to the material layers, as they break up the coherent volume perpendicular to **q** (Fig. 1). In operando BCDI measurements on P2-Na$_{0.66}$Ni$_{0.33}$Mn$_{0.66}$O$_2$ cathodes (Fig. 4, c), we have found multiple instances of screw dislocations nucleating perpendicular to the layers and peaks widening perpendicular to **q**. Notably, the Bragg peaks in P2-Na$_{0.66}$Ni$_{0.33}$Mn$_{0.66}$O$_2$ preserve the fringe and speckle structure and higher intensity throughout later stages of charging than in P2-Na$_{0.78}$Ni$_{0.23}$Mn$_{0.69}$O$_2$. Based on previous experiments in anion redox lithium-rich layered oxide cathodes[4], we attribute the differences in defect concentration at high voltage to a lower anion redox activity in Na$_{0.66}$Ni$_{0.33}$Mn$_{0.66}$O$_2$.

We evaluated strain in both materials by calculating the partial energy of strain in the [002] direction (Fig. 4, d) from BCDI reconstructions via $E_{p,002} = \frac{1}{2} Y_z \int |\varepsilon_{002}(\boldsymbol{r})|^2 d\boldsymbol{r}$, where $Y_z$ is the Young's modulus along [002] (estimated as 150 GPa[29]), $\varepsilon_{002}(\boldsymbol{r})$ is the measured strain as in Fig. 3, a-d, $\boldsymbol{r}$ is the spatial coordinate, and the integration is over the particle volume. The partial strain energy, $E_{p,002}$, in the P2-Na$_{0.78}$Ni$_{0.23}$Mn$_{0.69}$O$_2$ particle in Fig. 1-3 (thick solid blue line in Fig. 4, d) decreases as the domain boundary is formed at 28 mAh/g and increases again as the boundary resolves into a dislocation pair, supporting the notion that the domain boundary suppresses strain by storing energy. The strain energy density in both materials is 3-10 times lower in the initial charging stage than in Li-ion cells[4], staying below 0.1-0.2 pJ/um$^2$ during charging. The difference in the strain energy density suggests that open Na-ion paths in the P2 TMOs enable homogeneous sodium-ion diffusion in the particle[2]. The specific strain energy in Na$_{0.66}$Ni$_{0.33}$Mn$_{0.66}$O$_2$ up to ~70 mAHr/g (corresponding to the P2-O2 transition, see Fig. 4, e) is even lower (2-3 times) than in Na$_{0.78}$Ni$_{0.23}$Mn$_{0.69}$O$_2$, suggesting that the structural changes stall until the P2-O2 transition in Na$_{0.66}$Ni$_{0.33}$Mn$_{0.66}$O$_2$.

We reveal experimentally a formation mechanism for dislocation loops on the domain boundaries in sodium cathode nanoparticles in operando batteries. The role of external stress perpendicular to the layers ([001] direction) in the defect formation process offers a way to exploit texture and mesostructure in cathodes[29] for improving electrochemical performance. The observed lifetime of dislocation loops on the order of 30 mAHr/g demonstrates dislocation self-healing during charging - and further stresses the importance of operando over ex-situ measurements. Furthermore, our model and the observed planar strain signature within the loop offer a mechanism for the formation of stacking fault-type defects often observed ex-situ in layered cathode materials[1,30] through improper layer attachment left by the expanding loop. The observed probable retardation of ion diffusion from the central particle region by the domain boundary contrasts with theoretical considerations in Li-ion materials, which suggested that the boundary accelerates ion diffusion[14]. It also demonstrates a new and promising application of BCDI to track ion diffusion in 3D through improved structural resolution operando.

Finally, in all particles we reconstructed, for $Na_{0.78}Ni_{0.23}Mn_{0.69}O_2$ and $Na_{0.66}Ni_{0.33}Mn_{0.66}O_2$, we only found screw dislocations forming perpendicular to the layers. Preferential Bragg peak widening perpendicular to **q** in diffraction from individual particles corroborates preferential formation of defects perpendicular to the layers. Layer-perpendicular defect anisotropy directly contrasts edge dislocations oriented along the layers in LIB cathodes[4,11]. Unlike Li-ions, the Na-ions are larger than transition metal ions and cannot occupy the same crystal sites. The dislocations perpendicular to the layers can therefore serve as escape pathways for ions that cannot migrate through the TMO layers. The formation strain energy of the dislocations is proportional to their length, defined by the particle morphology, and the component of shear modulus along the Burgers vector. Both the edge and screw dislocations with the Burgers vector perpendicular to the layers depend on the same value of an orthotropic shear modulus; hence the ratios of the particle size parallel ($d_{||}$) and perpendicular ($d_{\perp}$) to the layers dictate the preferable dislocation formation. Consequently, in the plate-like morphology of P2-$Na_{0.66}Ni_{0.33}Mn_{0.66}O_2$ and P2-$Na_{0.78}Ni_{0.23}Mn_{0.69}O_2$ particles, screw dislocations normal to the layers are energetically more favourable. The morphology effect raises an intriguing possibility to design SIB cathode materials to minimize the number of defects or encourage their preferential orientation for controlling the speed of cation and anion diffusion.

# Methods

**Materials Synthesis.** P2-Na$_{0.78}$Ni$_{0.23}$Mn$_{0.69}$O$_2$ was synthesized using a titration technique as described previously[23]. 60 mL of a Na$_2$CO$_3$ solution was added dropwise to a 10 mL solution of Ni(NO$_3$)$_2$ and Mn(NO$_3$)$_2$ (Ni:Mn=1:3 molar ratio) where the CO$_3$:TM ratio was 1:1. The resulting solution was transferred to a 100 mL Teflon-lined stainless steel autoclave and aged at 80 °C for 12 h. The resulting Ni$_{025}$Mn$_{0.75}$CO$_3$ particles were mixed with a stoichiometric amount of Na$_2$CO$_3$, and the mixture was calcined at 900 °C for 12 h. P2-Na$_{0.66}$Ni$_{0.33}$Mn$_{0.66}$O$_2$ synthesized using a co-precipitation method. Stoichiometric amounts of precursors, Ni(NO$_3$)$_2$ and Mn(NO$_3$)$_2$ (Ni:Mn ratio=1:2 molar ratio) were dissolved in deionized water for a total concentration of 1M. The TM nitrate solution and a 0.2 M Na$_2$CO$_3$ aqueous solution were pumped separately into a reaction vessel to maintain a pH of 7.8. The obtained mixture was aged at 80 °C for 12 h. The resulting Ni$_{0.33}$Mn$_{0.66}$CO$_3$ was washed with deionized water and dried at 80 °C overnight. The Ni$_{0.33}$Mn$_{0.66}$CO$_3$ powder was mixed with a 5% excess stoichiometric ratio of Na$_2$CO$_3$ and calcined at 900 °C for 12 h.

**Electrochemical tests.** Cathode electrodes were prepared by mixing a slurry of 80 wt % active material with 10% acetylene black and 10% polyvinylidene fluoride (PVDF) with n-methyl-2- pyrrolidone as the solvent. The slurry was cast onto aluminium foil and dried at 80 °C under vacuum overnight. Na metal was used as the counter electrode with 1M NaPF$_6$ in propylene carbonate (PC) as the electrolyte and GF/F (Whatman) as the separator. Battery assembly was carried out in an MBraun glovebox (H$_2$O < 0.1 ppm). Modified *in situ* CR coin cells were used for CXDI experiments as previously described [5]. The voltage range for P2- $Na_{0.78}Ni_{0.23}Mn_{0.69}O_2$ was maintained between 2-4.5 V while the voltage range for P2- $Na_{0.66}Ni_{0.33}Mn_{0.66}O_2$ was maintained between 2.3-4.5 V. Electrochemical impedance spectroscopy (EIS) was carried out with 10 mV perturbation and AC frequencies from 100 kHz to 10 mHz. A SP-200 Biologic Potentiostat was used to measure impedance at different states of charge at a rate of C/10.

**Experimental details.** The experiments were conducted at the 34 ID-C beamline of the Advanced Photon Source (Argonne National Laboratory, ANL, USA). The batteries were mounted on standard sample holders manufactured using a 3D printer. A photon energy of 9 keV and sample-to-detector distances from 60 cm to 2 m were used in the experiments. Timepix (34ID) 2D detector with a pixel size of 55 µm × 55 µm was used. Fig. 4, e shows the electrochemical profiles measured simultaneously with the X-ray data on the in-situ cells.

**X-ray data collection and reconstruction procedure.** In all experiments rocking scans around a (002) Bragg peak, approximately 1° wide with 50–100 points and 0.5-2 second exposition, were collected. The scheme of the experiment and an example of individual diffraction patterns can be found in Supplementary Figure 7: BCDI scheme. During the analysis, the reconstruction procedure combined error-reduction (ER) alternating with hybrid input-output (HIO) algorithm. The diffraction data were binned by 2 along both dimensions of the detector before running reconstructions. Reconstruction without binning was also tested but produced worse results. Different iteration numbers between 410 and 1,000 were attempted, with the final number settled at 610. All attempts resulted in very similar reconstructions. We used an average of 5 results in this work, each being an average of 20 best reconstructions retrieved in a guided procedure developed in ref. 6 (8 generations, 40 population). The errors in Fig. 3 and Fig. 4 are calculated as a standard deviation between different reconstructions. The nanoparticle shape was found by averaging the amplitudes of the reconstructions at different voltages, assuming the nanoparticle shape change during charge is negligible, and applying a threshold of 10% to that average amplitude (see also ref. 5). The reconstructions were run using a GPU optimized code on multiple GeForce 1080 and 2080 graphics cards.

**Mechanical treatment.** For figures and detailed derivations, see Supplementary Note 3: Calculations of dislocation interaction energy and stress. Layered materials are approximated for simplicity as transverse isotropic with respect to the axis perpendicular to the layers (here $z$) with a compliance matrix

$$\begin{bmatrix}\varepsilon_{xx}\\ \varepsilon_{yy}\\ \varepsilon_{zz}\\ \varepsilon_{yz}\\ \varepsilon_{zx}\\ \varepsilon_{xy}\end{bmatrix}=\begin{bmatrix}\frac{1}{Y_p} & -\frac{\nu_p}{Y_p} & -\frac{\nu_{zp}}{Y_z} & 0 & 0 & 0\\ -\frac{\nu_p}{Y_p} & \frac{1}{Y_p} & -\frac{\nu_{zp}}{Y_z} & 0 & 0 & 0\\ -\frac{\nu_{pz}}{Y_p} & -\frac{\nu_{pz}}{Y_p} & \frac{1}{Y_z} & 0 & 0 & 0\\ 0 & 0 & 0 & \frac{1}{2G_{zp}} & 0 & 0\\ 0 & 0 & 0 & 0 & \frac{1}{2G_{zp}} & 0\\ 0 & 0 & 0 & 0 & 0 & \frac{1+\nu_p}{2Y_p}\end{bmatrix}\begin{bmatrix}\sigma_{xx}\\ \sigma_{yy}\\ \sigma_{zz}\\ \sigma_{yz}\\ \sigma_{zx}\\ \sigma_{xy}\end{bmatrix},$$

where $\varepsilon$ and $\sigma$ are the strain and stress tensors, $Y_p$ and $\nu_p$ are the Young modulus and Poisson ratio in the x-y symmetry plane, $Y_z$ and $\nu_{zp}$ in the z-direction, and $G_{zp}$ is the shear modulus in the z direction. Here $\frac{\nu_{pz}}{Y_p}=\frac{\nu_{zp}}{Y_z}$. Critical radius $r_c$ and the activation energy $E_c$ per unit length for the homogeneous formation of a dislocation pair of two screw dislocations of opposite chirality in the z direction given shear stress $\sigma_{zp}$ are derived to be

$$r_c = G_{zp}b/\pi\sigma_{zp}, \qquad E_c = G_{zp}b^2(2\ln(r_c/r_0)-1+\ln 2)/2\pi,$$

where $b$ is the Burgers vector.

Shear stress needed to push the dislocations to the particle surface despite their attraction is estimated as

$$\sigma_{zp} \approx G_{zp}b/2\pi r_c - G_{zp}b/2\pi r_{surf},$$

where $r_c$ is the initial distance between dislocations after their formation and $r_{surf}$ is the distance from the dislocations to the surface, which produces an image force pulling each dislocation to the surface due to a missing part of the displacement/strain field.

## Data availability
The data that support the plots within this paper and other findings of this study are available from the corresponding authors upon reasonable request. The X-ray imaging data are deposited at Sector 34-ID-C of the Advanced Photon Source.

## Acknowledgements

The work at Cornell was supported by the National Science Foundation under Grand No. (CAREER DMR 1944907). The work at UC San Diego was supported by the National Science Foundation (NSF) under Award Number DMR1608968. The SEM analysis in this work was performed at the San Diego Nanotechnology Infrastructure (SDNI), a member of the National Nanotechnology Coordinated Infrastructure, which is supported by the National Science Foundation (grant ECCS1542148). This research used resources of the Advanced Photon Source, a U.S. Department of Energy (DOE) Office of Science User Facility, operated for the DOE Office of Science by Argonne National Laboratory under Contract No. DE-AC02- 06CH11357.

## Author contributions
O.Yu.G., H.H., D.S., D.W., R.B., Z.W., W.C., J.M, R.H., and A.S. conducted the coherent X-ray measurements; H.H., M. Z., Y.S.M. synthesized the cathode materials, performed the scanning electron microscopy, EIS, and assembled the batteries; O.Yu.G. performed data reduction, analysis and mechanical interpretation, with contributions from H. H., D.S.M.,Y.S.M., and A.S.; O.Yu.G., H.H., and A.S. wrote the paper. All authors contributed to discussions and commented on the manuscript.

## Competing Interests statement
The authors declare no competing interests.

# Figures

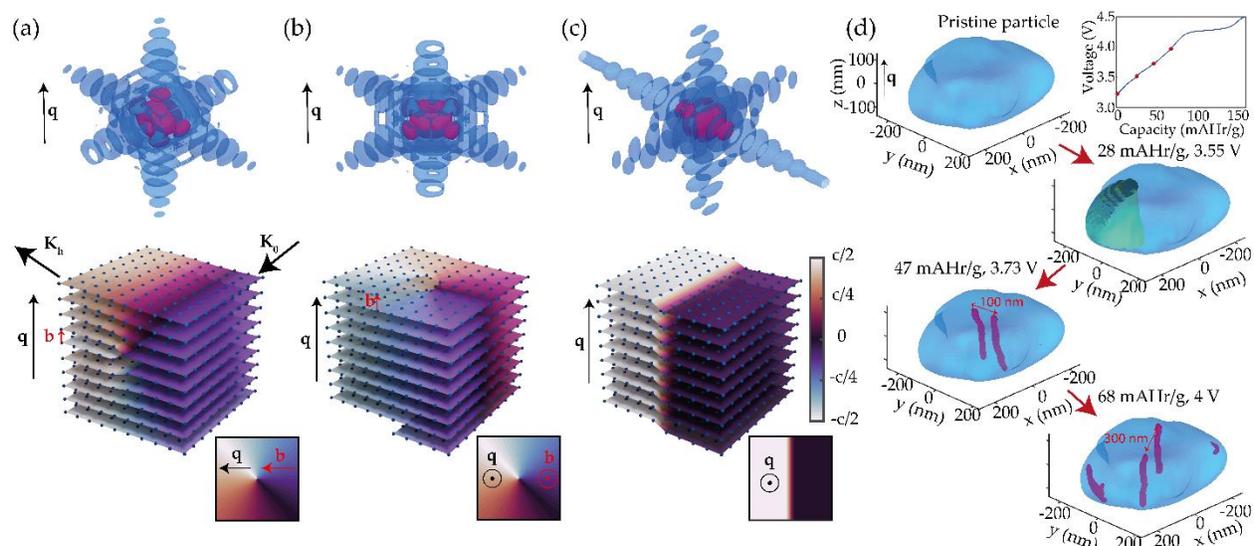

Figure 1: Operando Bragg Coherent Diffractive Imaging (BCDI). Crystal defects (edge dislocation **a**, screw dislocation **b**, antiphase domain boundary **c**) generate characteristic 3D reciprocal space patterns (**top**) in coherent X-ray scattering. The phase retrieval algorithm allows inversion of the diffraction profiles into 3D electron density and atomic displacements of the scattering planes (**bottom**). Here $K_0, K_h, q$ are the incident and scattered wavevectors and a scattering vector respectively, $c$ is the lattice constant perpendicular to the layers. An analysis of the dimensionality and orientation of the singularities in atomic displacements allows identification of the type of defect. **d,** Isosurface rendering (blue transparent) of a P2-$Na_{0.78}Ni_{0.23}Mn_{0.69}O_2$ particle measured during charge. Pristine particle displays no observable defects (top), a domain boundary develops at 3.55V (green, boundary drawn as isosurface at $c/4$ displacement level), two dislocations form (red/magenta) at 3.73 V, recognized based on **a-c**, and the dislocations move apart 4 V. Electrochemical data is shown in the top right inset, red points show the x-ray measurements.

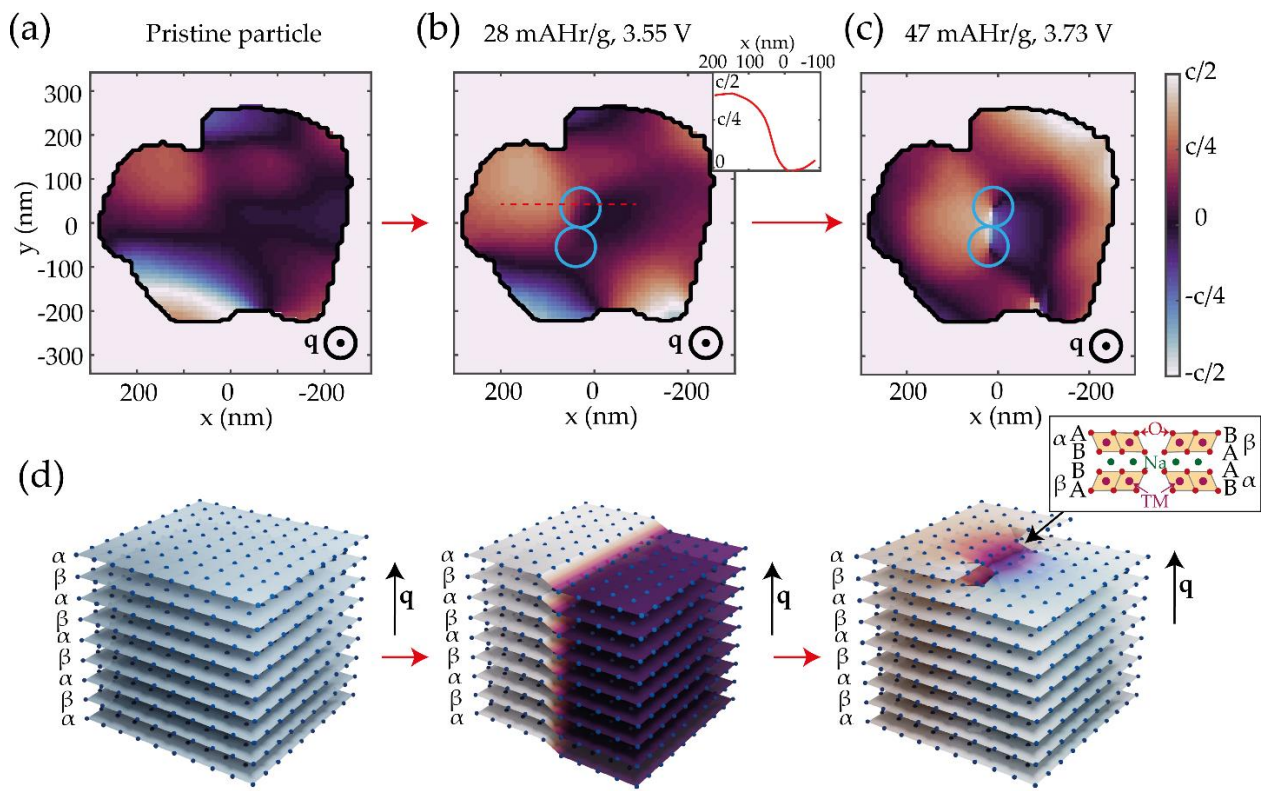

Figure 2: Dislocation pair nucleation on a domain border in P2-Na$_{0.78}$Ni$_{0.23}$Mn$_{0.69}$O$_2$. **a, b, c,** Comparison of displacement cross sections at different points during charging. Appearance of 2 screw dislocations on the former domain boundary is visible (blue circles). Dislocations have antiparallel Burgers vectors (left-handed and right-handed screw dislocation). **d**, Schematic representation of a glissile dislocation loop nucleation on the domain boundary. Note the A to B connection after nucleation (inset).

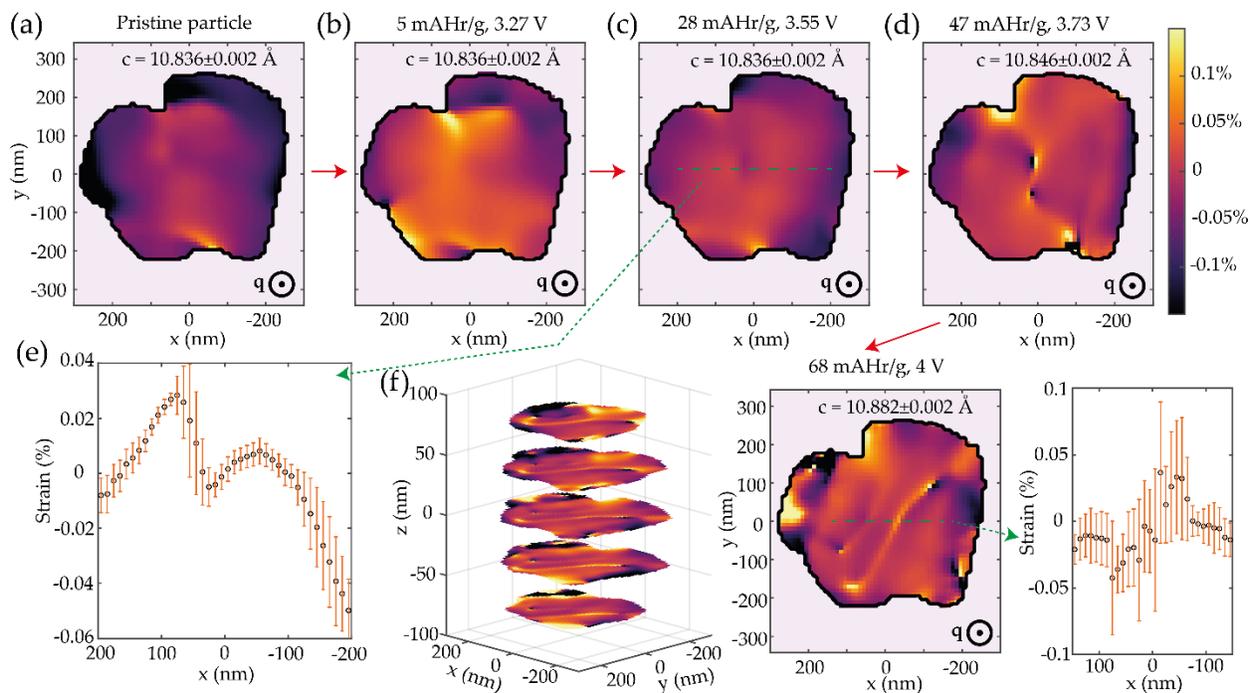

**Figure 3: Evolution of the strain. a-d,** Cross sections of the strain distribution in a P2-$Na_{0.78}Ni_{0.23}Mn_{0.69}O_2$ crystal particle at different stages of battery charging. The average lattice constant c around which the strain is calculated is also shown. **e,** Linear cross section of the strain, averaged along **q**, along the green line in **c**, showing a 0.02% drop in lattice constant in the particle centre. Error is calculated as a standard deviation between different reconstructions. **f,** Strain distribution in the particle in the middle of the phase transition, showing a "strain wall" between the dislocations.

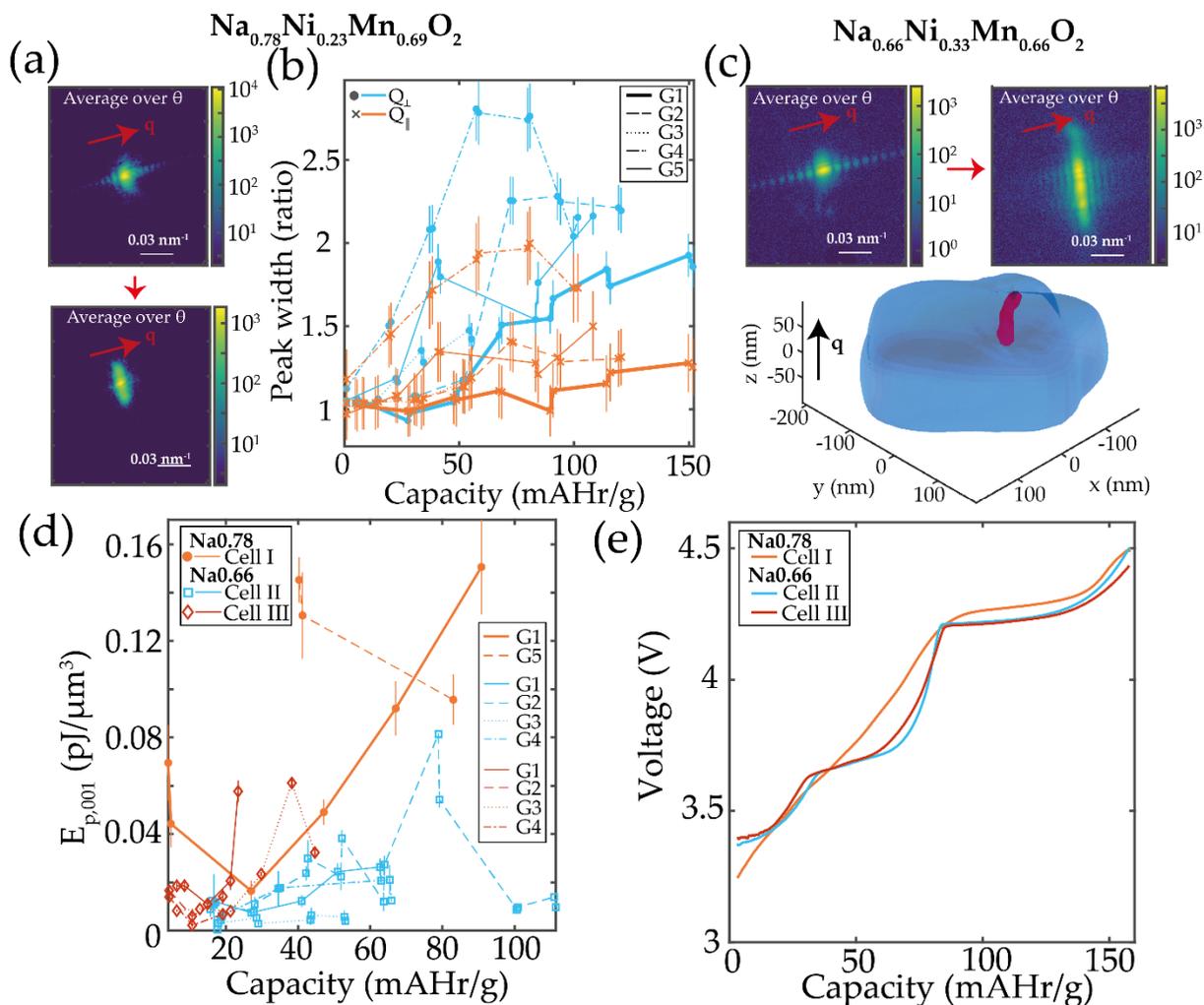

Figure 4: Expanded analysis over many cathode particles. **a,** Example of evolution of projected (averaged over the reflection angle θ) Bragg peaks for P2-Na$_{0.78}$Ni$_{0.23}$Mn$_{0.69}$O$_2$. Top - pristine, bottom - during charging. **b,** Evolution of the peak width for multiple grains, along (x markers, orange) and perpendicular (o markers, blue) to the **q** direction. Grain shown in previous figures is G1. Error is calculated as an error on width (through kurtosis). Includes **c,** Example of evolution of projected (averaged over the reflection angle θ) Bragg peaks for P2-Na$_{0.66}$Ni$_{0.33}$Mn$_{0.66}$O$_2$. Left - pristine, right - during charging. Example of a reconstructed particle (blue) with a dislocation (red) for P2-Na$_{0.66}$Ni$_{0.33}$Mn$_{0.66}$O$_2$. **d,** Specific strain energy for different particles of both materials in different cells in the process of charging. Errors are calculated as a standard deviation between strain energy calculated for different particle surface assumptions (10%, 15%, 20% of the maximum amplitude). **e,** Voltage curves of different measured cells.

# Supplementary Information for

# Interaction and transformation of metastable defects in intercalation materials

O. Yu. Gorobtsov[1], H. Hirsh[2], M. Zhang[2], D. Sheyfer[3], S. D. Matson[1], D. Weinstock[1], R. Bouck[1], Z. Wang[1], W. Cha[3], J. Maser[3], R. Harder[3], Y. Sh. Meng[2], A. Singer[1]

1 – Materials Science and Engineering Department, Cornell University, Ithaca, NY 14853, USA
2 – Department of NanoEngineering, University of California, San Diego, La Jolla, California, 92093, USA
3 - Argonne National Laboratory, Argonne, Illinois 60439, USA

**This PDF file includes:**

**Supplementary Note 1: Electron microscopy and XRD**

For the two cathode materials P2-$Na_{0.78}Ni_{0.23}Mn_{0.69}O_2$ and P2-$Na_{2/3}Ni_{1/3}Mn_{2/3}O_2$ studied in this paper, we performed x-ray powder diffraction (XRD) to verify crystal structure and determine optimal peaks for the BCDI measurements. XRD results are shown in Supplementary Figure 1 and Supplementary Figure 2. We performed BCDI measurements on the (002) peak, as [002] direction is perpendicular to the layers of the material and the (002) peak has the highest intensity. We performed scanning electron microscopy on the pristine synthesized cathode particles to determine their typical morphology. The particles sizes vary from several hundred nanometers to several microns. The particles demonstrate a pronounced plate-like morphology.

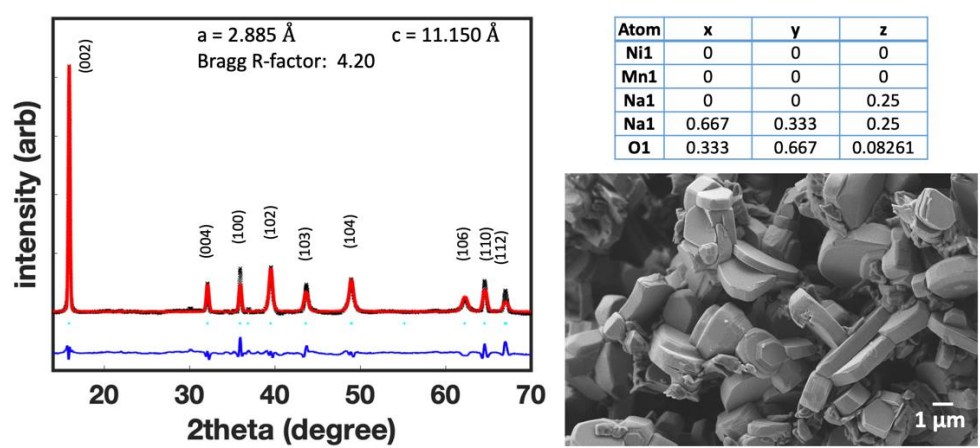

**Supplementary Figure 1.** Refined XRD data of P2-$Na_{0.78}Ni_{0.23}Mn_{0.69}O_2$ with refinement parameters and an SEM image of the pristine powder.

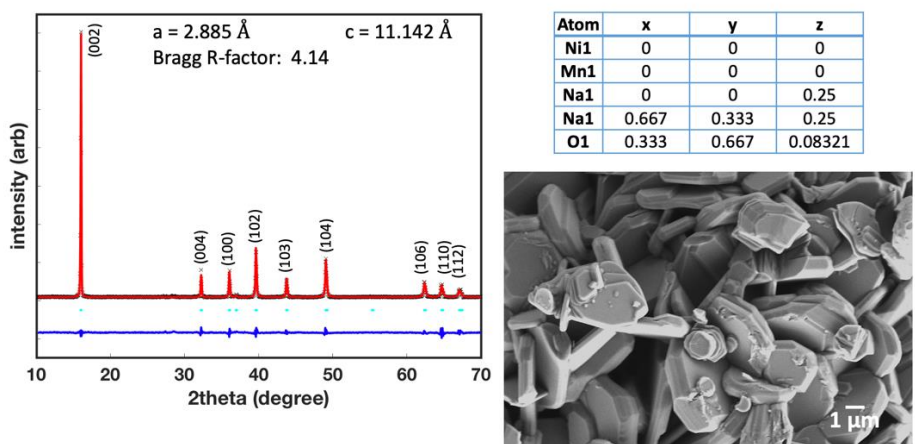

**Supplementary Figure 2.** Refined XRD data of P2-$Na_{2/3}Ni_{1/3}Mn_{2/3}O_2$ with refinement parameters and an SEM image of the pristine powder.

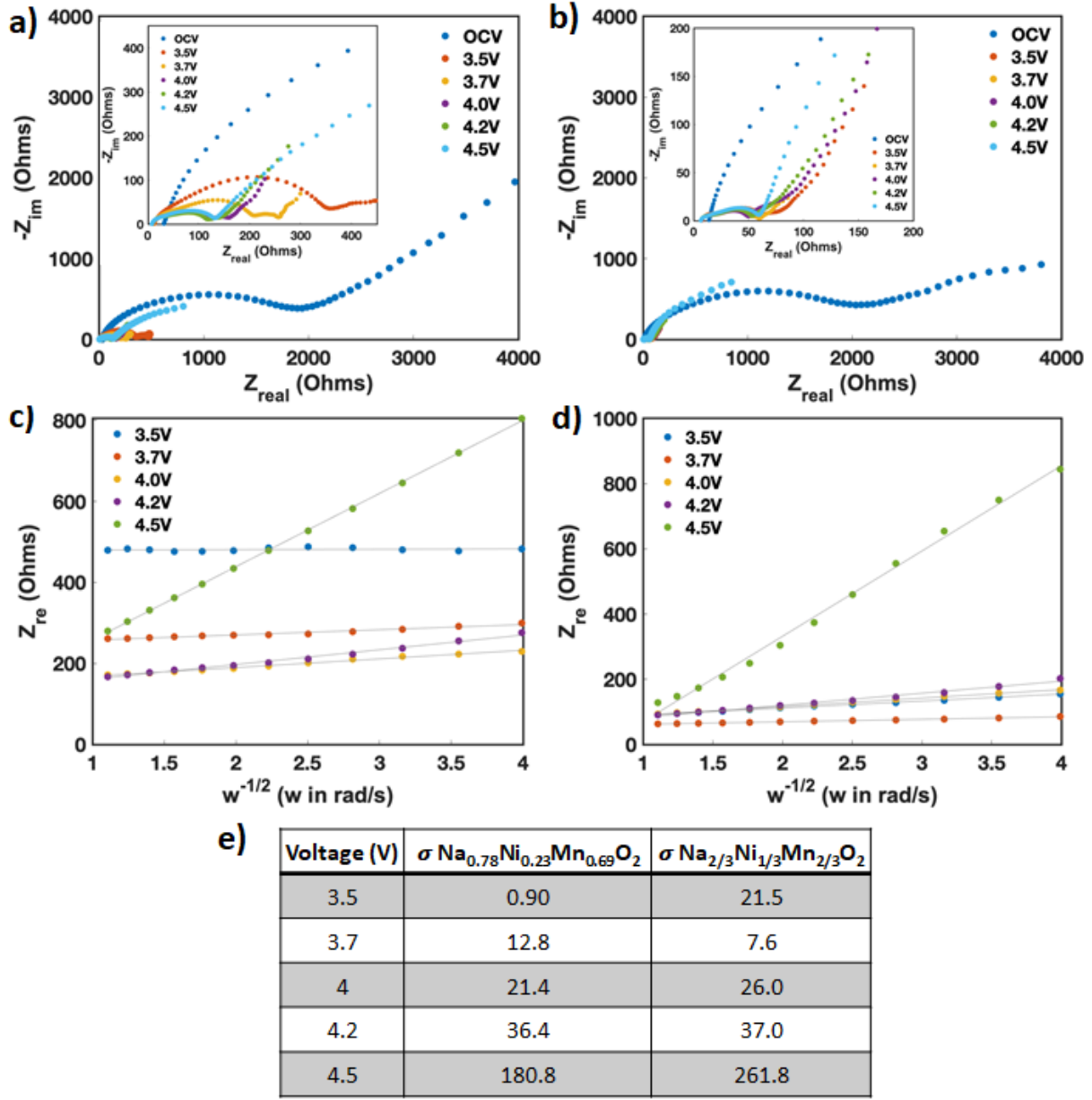

**Supplementary Figure 3. a, b** Nyquist plots at different states of charge for $Na_{0.78}Ni_{0.23}Mn_{0.69}O_2$ and $Na_{2/3}Ni_{1/3}Mn_{2/3}O_2$ during the first charge at a rate of C/10 and their respective **c, d** plots of $Z_{re}$ vs. the inverse square root of angular frequency. The slope $\sigma$ represents the Warburg coefficient $Z_{re} \propto \sigma\omega^{-1/2}$ which has the following relationship with the bulk diffusion constant: $D \propto \frac{1}{\sigma^2}$ [1]. **e,** $\sigma$ at different states of charge is listed in the table.

**Supplementary Note 2: Calculations of dislocation interaction energy and stress**

Application of external stress to the material can cause formation of defects in the material. Nucleation Approximating in the interest of simplicity layered materials as transverse isotropic with respect to the axis perpendicular to the layers (z for the purposes of this Supplementary), the compliance matrix is expressed as [2]

$$\begin{bmatrix}\varepsilon_{xx}\\ \varepsilon_{yy}\\ \varepsilon_{zz}\\ \varepsilon_{yz}\\ \varepsilon_{zx}\\ \varepsilon_{xy}\end{bmatrix}=\begin{bmatrix}\frac{1}{Y_p} & -\frac{\nu_p}{Y_p} & -\frac{\nu_{zp}}{Y_z} & 0 & 0 & 0\\ -\frac{\nu_p}{Y_p} & \frac{1}{Y_p} & -\frac{\nu_{zp}}{Y_z} & 0 & 0 & 0\\ -\frac{\nu_{pz}}{Y_p} & -\frac{\nu_{pz}}{Y_p} & \frac{1}{Y_z} & 0 & 0 & 0\\ 0 & 0 & 0 & \frac{1}{2G_{zp}} & 0 & 0\\ 0 & 0 & 0 & 0 & \frac{1}{2G_{zp}} & 0\\ 0 & 0 & 0 & 0 & 0 & \frac{1+\nu_p}{2Y_p}\end{bmatrix}\begin{bmatrix}\sigma_{xx}\\ \sigma_{yy}\\ \sigma_{zz}\\ \sigma_{yz}\\ \sigma_{zx}\\ \sigma_{xy}\end{bmatrix}, \quad (1)$$

where $\boldsymbol{\varepsilon}$ and $\boldsymbol{\sigma}$ are the strain[1] and stress tensors, $Y_p$ and $\nu_p$ are the Young modulus and Poisson ratio in the x-y symmetry plane, $Y_z$ and $\nu_{zp}$ in the z-direction, and $G_{zp}$ is the shear modulus in the z direction. Here $\frac{\nu_{pz}}{Y_p}=\frac{\nu_{zp}}{Y_z}$.

Consider a pair of parallel screw dislocations (Supplementary Figure 4, a) oriented along z. The radial and tangential components of the interaction force per unit length between them are well known [3]

$$F_r = G_{zp}b^2/2\pi r, \quad F_\theta = 0, \quad (2)$$

where $\boldsymbol{b}$ is the Burgers vector and $\boldsymbol{r}$ is the distance vector. $F_r$ is attractive for oppositely oriented screws. Elastic interaction energy per unit length, which is essentially an energy of introducing a second dislocation, is then

$$E_{int} = \int_{r_0}^{r} G_{zp}b^2/2\pi r \, dr = G_{zp}b^2 \ln(r/r_0)/2\pi, (3)$$

where $r_0$ is the core dislocation radius. Considering the elastic energy per unit length of a single screw dislocation is $E_{el}(screw) = G_{zp}b^2 \ln(R/r_0)/4\pi$ [3], where R is the outer radius, and that the stress fields of the two screw dislocations of opposite chirality will roughly cancel each other starting at the distances of $\approx 2r$ from the pair, the full elastic energy is then

$$E_{tot} = G_{zp}b^2(2\ln(r/r_0) + \ln 2)/2\pi. \quad (4)$$

---

[1] Note that the shear strain here is not the engineering shear strain.

Imagine now that such a dislocation pair is nucleated homogeneously under an applied shear stress $\sigma_{zp}$. The work per unit length done by the applied stress is then $A = r\sigma_{zp}b$, and the associated increase in energy per unit length is

$$E = G_{zp}b^2(2\ln(r/r_0) + \ln 2)/2\pi - r\sigma_{zp}b, \qquad (5)$$

from which the critical radius $r_c$, at which the dislocation pair will be stable, and the corresponding activation energy per unit length $E_c$ are determined by differentiation as

$$r_c = G_{zp}b/\pi\sigma_{zp}, \qquad E_c = G_{zp}b^2(2\ln(r_c/r_0) - 1 + \ln 2)/2\pi. \quad (6)$$

The distance between dislocations and therefore critical radius in a nanocrystal is limited by the size of the nanocrystal, establishing a lower bound on the necessary shear stress. The plate-like nanocrystal shape with the size perpendicular to the layers approximately 3-5 times smaller than parallel makes formation of the pairs perpendicular to the layers significantly more energy efficient. Using the observed distance between the dislocations after formation $r_c = 100\ nm$, the shear modulus estimate $G_{zp} \sim 50\ GPa$ [4] [5] [6], and the Burgers vector b=0.5 nm we find the minimum stress necessary for the pair formation as $\sigma_{zp} \approx 80$ MPa. Core radius of a dislocation can be estimated as $r_0 \approx l_p/(1-v) \approx 5\ nm$ [7], where $l_p$ is the lattice period in the layer plane and $v$ is the Poisson ratio, and the activation energy per unit length is then $E_c \approx 1.1 \cdot 10^{-8} \frac{J}{m}$.

    A pair of screw dislocations with opposite Burgers vectors experiences an attractive force. Therefore, the drift of the pair away from each other to the borders of the particle observed in the experiment requires a continuous external stress, which can be estimated as $\sigma_{zp} \approx G_{zp}b/2\pi r_c > 40\ MPa$. Note that a screw dislocation close to the particle surface experiences an image force (see Supplementary Figure 4, b) due to a missing part of displacement/strain field. The image force can be treated as an attractive force between a screw dislocation and its mirror image. The initial stress required to push the dislocation pair apart is therefore weakened ~2 times (initial distance from the dislocations to the surface is ~200 nm) and can be estimated as $\sigma_{zp} \approx 20\ MPa$.

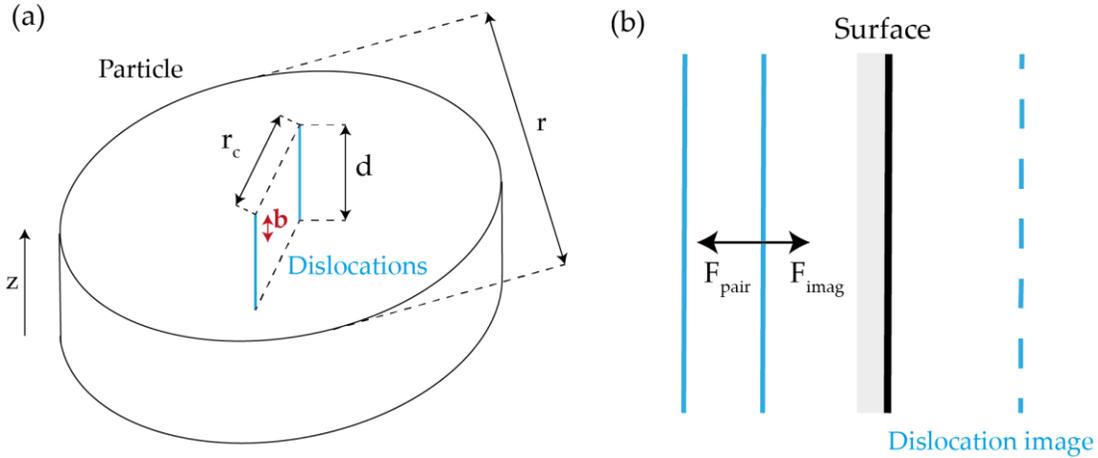

**Supplementary Figure 4.** Schematic representation of the variables in the Supplementary Note 2. **a**, Grain scheme. **b**, Image dislocation.

## Supplementary Note 3: Additional illustrations of the domain boundary

During the charging process of the P2-Na$_{0.78}$Ni$_{0.23}$Mn$_{0.69}$O$_2$ cell, a domain of a homogeneous high displacement develops in the particle. The domain is visible throughout the whole particle in 3D (Supplementary Figure 5), with the border perpendicular to the layer direction, as expected for an antiphase or out-of-phase domain. As the P2-Na$_{0.78}$Ni$_{0.23}$Mn$_{0.69}$O$_2$ cathode is further charged, the nucleation of a pair of line defects happens clearly on the boundary of the former domain. While the domain boundary itself disappears, the location of the defect pair clearly matches the boundary location (Supplementary Figure 6).

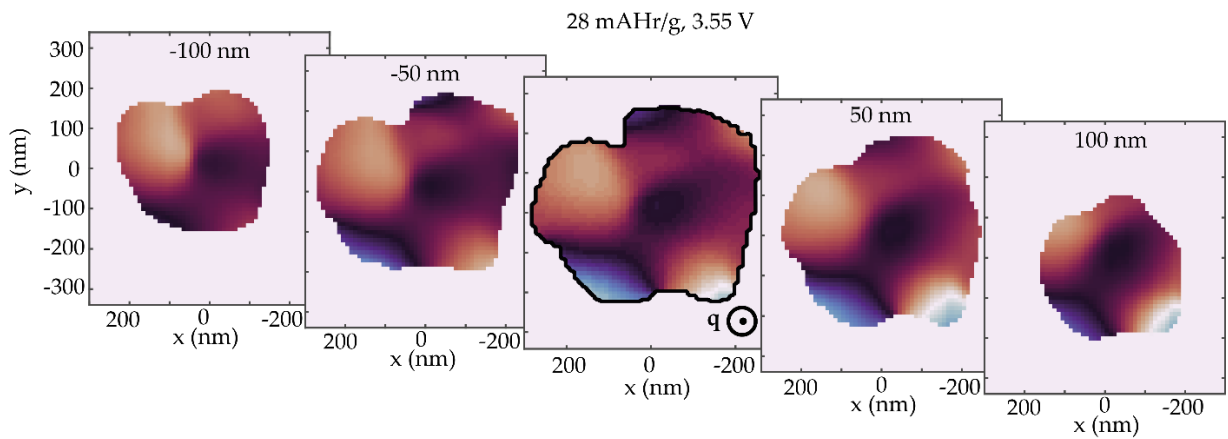

**Supplementary Figure 5.** Cross sections of displacement (similar to Figure 2) through the domain boundary at different *z* positions in the nanocrystal.

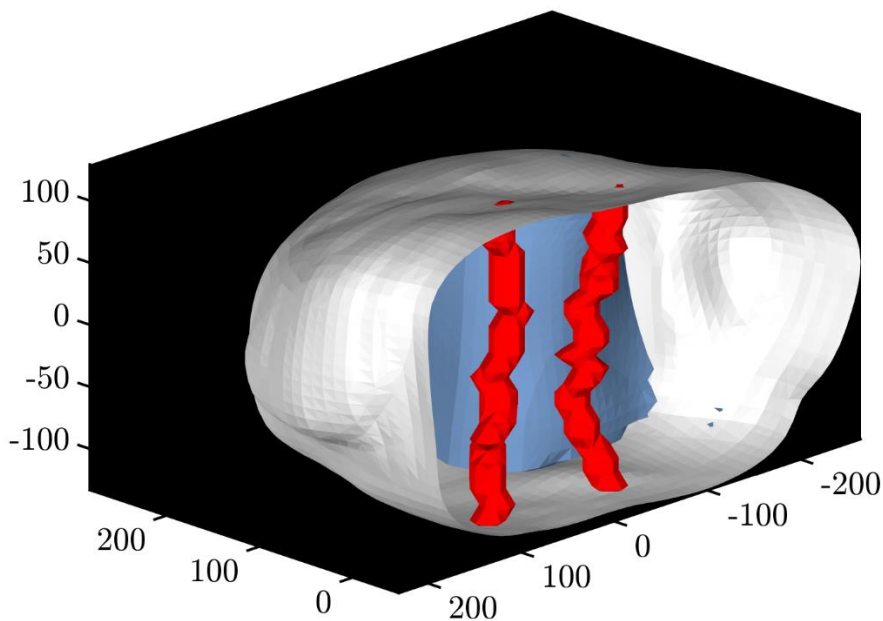

**Supplementary Figure 6.** Relative position of the domain boundary and the dislocation pair it generates. Blue - domain boundary (isosurface at $\pi/2$; red - dislocations. Merged from two subsequent measurements.

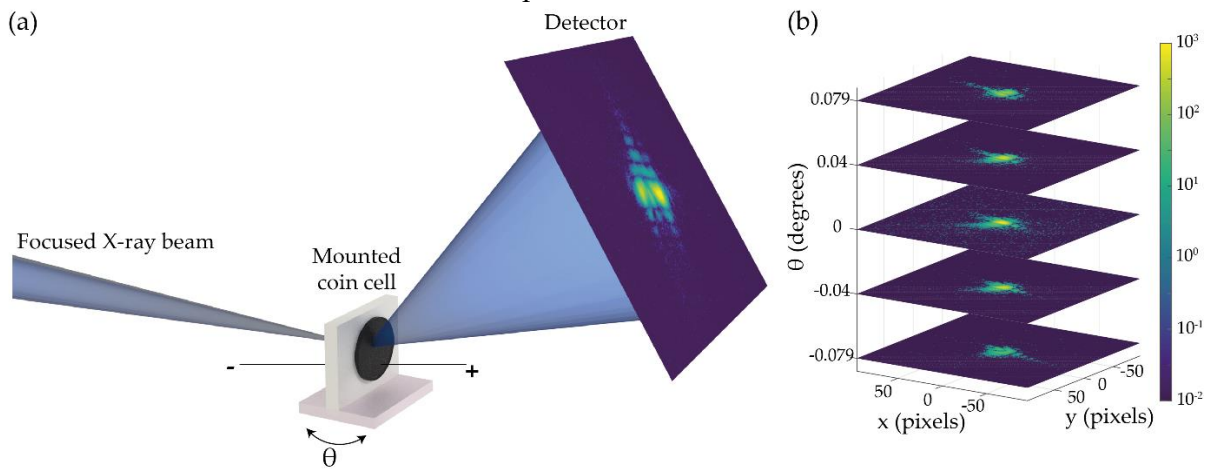

**Supplementary Figure 7: BCDI scheme. a**, Charging coin cell mounted on a sample stage. Incident focused X-rays scatter from the particles in the cathode, and the scattering is observed on the detector. The sample stage can be rotated in the scattering plane. **b**, 3D diffraction pattern is collected by scanning the rocking curve and then used to reconstruct the 3D displacement field.